# Where do free-ranging dogs rest? A population level study reveals hidden patterns in resting site choice.

Sourabh Biswas[1], Kalyan Ghosh[2], Kaushikee Sarkar[3] & Anindita Bhadra[1*]

sb18rs107@iiserkol.ac.in; kalyanghosh1999@gmail.com; kaushikeesarkar318@gmail.com; abhadra@iiserkol.ac.in

[1]Behaviour and Ecology Lab, Department of Biological Sciences, Indian Institute of Science Education and Research Kolkata, India.

[2]Department of Zoology, University of Gour Banga, India

[3]Department of Environmental Science, University of Calcutta, India.

**Abstract:**

Rest is essential for animals to retain energy and alleviate stress. However, free-ranging dogs in human-dominated areas encounter obstacles such as noise, pollution, limited food sources, and anthropogenic disturbance while resting. Since free-ranging dogs have survived as a population in India, as in many other parts of the Global South for centuries, they provide a unique opportunity to study adaptation of animals to the human-dominated urban landscape. From 2019 to 2022, we explored factors impacting resting behaviour and site preferences in 66 dog groups (284 dogs) in three states of India. Observations were carried out in eight three-hour time windows to capture fluctuations throughout the day, noting resting place features and behavioural states. Seven physical parameters of the resting sites, along with the biological factors like mating and pup-rearing and time of day affected their choice of resting sites. The frequency-rank distribution of the various unique combinations in which the seven parameters were selected by the dogs followed a Power law distribution, which suggests underlying biological reasons for the observed preferences. Further analysis revealed that three of these parameters showed maximum consistency of choice in terms of the sub-parameters selected, and this explained 30% of the observations. We could thus conclude that the dogs prefer to rest close to their resource sites within the territory, at a place that enabled maximum visibility of the surroundings. When they wanted to sleep, they chose such sites in the core of the territory. At other times, they chose such sites away from the core, and were less restive, thus remaining at a state that allowed for immediate response, in case of intrusion or threat. They generally avoided anthropogenic disturbance for sleeping, and preferred areas with shade. Our findings show that dogs chose resting positions depending on resource defence, foraging effectiveness, thermoregulation, and territorial defence. Incorporating these aspects into urban management plans can promote human-dog cooperation and reduce situations of conflict. We envisage more inclusive urban areas in the future, that can allow for co-existence of the humans and their oldest companions in the commensal relationship that has been maintained for hundreds of generations of dogs in this part of the world.

**Introduction:**

Animals have a physiological requirement for resting, which includes sleep and other energy conserving behaviours, with a high degree of variation from species to species. Among mammals, such variation has been studied with respect to posture and site of resting (Purcell et al., 2009; Taylor & Buskirk, 1994) and time spent in resting (Jucá et al., 2020; Society, 2017). Mammals show huge variation in their resting needs, which can range from 3 to 20 hours of a day (Lutermann et al., 2010). Mammals have been shown to sleep in a wide variety of locations, including open areas in vegetation, natural shelters like tree cavities, constructed shelters like nests, dens, and warrens, as well as in man-made constructions (Kunz & Lumsden,

2003; Fisher & Wiebe, 2006; Herr et al., 2010; Schneider et al., 2013). The decision of where to sleep or rest has a significant impact on an animal's health and survival (Brown et al., 2014).

Animals' choice of sleeping place is influenced by body size, social structure, thermoregulation requirements, and protection from the weather (Anderson, 2000; Chivers, 1974) Reproductive status also plays a role, with individuals choosing safer locations when young are present (Henner et al., 2004; Sen Majumder et al., 2016). The placement of food supply can also impact sleeping location, with animals choosing central locations or those close to recent eating spots (Chapman et al., 1989; Smith et al., 2007). Insects and other arthropods can also influence resting place selection (Feilen & Marshall, 2017; Nunn & Heymann, 2005).

As the world becomes more urbanized, artificial infrastructure is replacing natural ecosystems, leading to a decrease in natural resources and increased anthropogenic disturbance levels for animals (Hamer & McDonnell, 2010; Lowry et al., 2013). This has resulted in a general loss of biodiversity, known as "biotic homogenization" (McKinney, 2006; Mckinney, 2008). Despite this, many species have adapted and thrived in urban environments and are referred to as urban "adapters" or "exploiters" (McKinney, 2006). Urban adapted animals require resting to survive and maintain their well-being in small green spaces of cities. The stressors they face include noise, pollution, predators, limited access to resources, and constant disturbance. Resting conserves energy, reduces stress, repairs damage, and helps them avoid predators. However, finding a safe place to rest can be challenging.

Canids, including meso-carnivores like coyotes, gray wolves, dingoes, and red foxes, have a vast geographic range and inhabit various environments, from untouched woods to urban areas, but Arctic foxes and dholes avoid human settlements (Jackson et al., 2014; Mech, 1989;

Meriggi et al., 1991). Gray wolves underwent domestication, leading to the domesticated dog (Clutton- Brock, 1995). Though domesticated dogs are widely studied, free-ranging dogs form the majority of the world's dog population, being widely distributed primarily in the Global South.

Free-ranging dogs in India are scavengers found in human settlements and reproduce near humans (Vanak et al., 2009; Spotte, 2012). They exhibit complex social behaviour, forming social groups, engaging in territorial defence, maternal and alloparental care as well as parent-offspring conflict; (Paul et al., 2014). They interact with humans, with whom they share their habitats and engage in both positive and negative social interactions (Sen Majumder et al., 2014). These social bonds are crucial for their survival in urban and suburban environments where they rely on humans for food and shelter. Free-ranging dogs also have the ability to understand human gestures and vocalizations (Coppinger et al., 2017), but their social behaviour can vary depending on environmental conditions and population density. These dogs have high behavioural plasticity and prefer to give birth and rear pups near humans, despite high pup mortality due to human interference, making them an excellent model system to understand the spatial distribution and habitat use of urban-adapted animals (Sen Majumder et al., 2016). While pet dogs live entirely depending on people for their care and provisioning, free-ranging dogs mostly live on anthropogenic food waste since they are not directly supervised by people (Serpell, 1995). They can be a good model system for studying urban adaptation in mammals due to their ubiquitous presence in urban areas and their ability to adapt to the environmental challenges of living in such areas. Studying their behaviour, physiology, and genetics can provide insights into how mammals can adapt to urban environments. In addition, understanding the behaviour and ecology of free-ranging dogs can help mitigate human-dog and human-animal conflicts in urban areas. By identifying factors that contribute to conflict, such as resource availability and spatial use patterns, we can develop effective management strategies to reduce these conflicts. For instance, allowing free-ranging dogs suitable resting spaces can lower their stress levels and reduce the possibility that they will get into fights with people or other animals. This can facilitate coexistence and decrease the risks brought on by their presence, such as the spread of zoonotic illnesses. Overall, research on free-ranging dogs can assist improve cooperation between people and animals in cities by shedding light on how to recognize and manage urban wildlife populations.

For dogs, sleeping is a necessary behaviour because it allows them to store energy and recover from exercise. We carried out a long term study in different areas of India, to understand the factors influencing free-ranging dogs' resting behaviour and explore their preferences for resting sites. We expected them to rest away from anthropogenic disturbance, in places that provided protection from the weather and intruders, and near food sources, within their territories.

**Methodology:**

**Sampling:** Free-ranging dog groups (N = 66) were selected through surveys between 2019 to 2022 from West Bengal, Maharashtra and Rajasthan, India, in four different seasons: pre-mating, mating, pup-rearing and rest of the year. Of these, 54 groups were from various areas in West Bengal. While there were six groups each from Banswara, Rajasthan and Pune, Maharashtra (Table S1, Figure S1). The localities were chosen arbitrarily and based on convenience, in business or residential areas or both. It should be noted that part of this work was carried out during the pandemic, so selection of areas was regulated by travel restrictions.

In order to carry out observations, a human day was divided into eight time slots of three hours each (0000-0300hr, 0300-0600hr, 0600-0900hr, 0900-1200hr, 1200-1500hr, 1500-1800hr, 1800-2100hr, 2100-2400hr; see (Banerjee & Bhadra, 2022). Observations in each time slot were carried out six times per group per season. The observer walked along a predetermined route and on sighting a dog or cluster of dogs (dogs present within one body length of each other) resting, the observer recorded the behaviour, sex, life stage of the dog or dogs during resting state. The observer also noted various physical properties of the resting place, i.e., distance from food source and water, height and level of the site, insect presence, position of the dog within its territory, nature of resource available, level of anthropogenic disturbance, position and surface of resting site, visibility from the resting site, presence of shade, lighting condition, temperature and humidity using a customized application, 'dog-e' built for the android platform (see Table S2 for details). The nature of the resources in the territory and group size were also recorded accordingly. Henceforth, these have been denoted as parameters of the resting sites. A total of twelve physical parameters were recorded for each observation, to describe the resting site characteristics.

**Statistical analysis:** All the statistical analysis was carried out in R studio (version 4.2.0). We used an ordinal logistic mixed model (OLMR) with "ordinal" package (Haubo et al., 2022) in R to understand the impact of resting site characteristics and climatic condition on the resting behaviour of dogs. For the OLMR analysis, behaviour category was taken as the response variable while the resting site characteristics (Table S2) and climatic condition (temperature and season) were taken as predictor variables, keeping the identity of the dogs as a random factor. We grouped the behaviours (Table S3) into three categories – slightly relaxed; moderately relaxed and extremely relaxed in the context of energy conservation in dogs (see

Banerjee & Bhadra, 2022 for details). To understand the effect of resting site characteristics and climatic factors on the proportion of dogs resting in a group, we used "glmmTMB" package (Brooks et al., 2017) to perform GLMM with a beta distribution for resting proportion (number of dogs resting in the territory divided by the total number of dogs in the group), considering resting proportion as the response variable, resting site characteristics, temperature and season as predictor variables and dog identity as a random variable. All the final models were selected based on the lowest AIC value. The null and full models were compared for all the models. Dispersion of the models and residual diagnostics were checked using the "performance" package of R (Lüdecke et al., 2021).

We estimated the frequencies at which resting sites with each unique combination of parameters were selected in the population and fitted the distribution to a Power Law using the "poweRlaw" package (Gillespie, 2015)This was done for the entire data set, using the 7 parameters for which the earlier analysis showed significant variation (Table S4). Since different individuals had been sighted for a variable number of times, the data was normalized to proportion of selection by a unique combination of characters. Further, the same was checked for each category of resting behaviour (slightly relaxed, moderately relaxed, extremely relaxed).

Kruskal-Wallis rank sum test and Wilcoxon signed rank test were carried out to compare the parameters of the resting sites. We have used Dunn's test with Bonferrani correction for post hoc pair wise comparison. The alpha level was kept at 0.05 throughout the analysis.

**Results:**

We collected data on 6047 sightings of free-ranging dogs. Dog groups were sampled repeatedly, and sightings at the group level ranged from 2 to 19. Individual dogs were sighted 1 to 83 times. Of these, around 19% dogs were sighted as "slightly relaxed", 43% dogs were "moderately relaxed" and 38% of dogs were "extremely relaxed". 954 dogs were sighted as resting alone, while 5093 dogs were sighted to be resting in clusters of two or more.

**Quality of Rest**

Resting behaviour of dogs was found to be affected by a multitude of factors – the distance of the resources from the resting site, anthropogenic disturbance, position of the resting site within the territory, resting surface, shade, temperature, light condition of the resting site, visibility from the resting site and season of the year.

Ordinal mixed logistic regression (OMLR) analysis revealed that dogs tend to be more restive when they are not close to food sources '>10 m' (Z= 5.53, p < 0.001) (Table S4). They are more relaxed when they are resting in an area with moderate disturbance (Z= 2.01, p < 0.05), in both low (Z= 4.88, p < 0.001) and moderate (Z= 5.96, p < 0.01) light conditions (Table S4). Dogs were found to be less restive in both intermediate (Z= -2.04, p < 0.05) location and in the edge (Z= -8.01, p < 0.001) of the territory and were more relaxed when they rested deep within their territory. Resting in the vicinity of humans (Z= -9.59, p < 0.001) or on the road (Z= -1.90, p < 0.05) made them less relaxed. They appear to be more relaxed in shades (Z= 10.03, p < 0.001) and less relaxed with the increase in temperature (Z= -14.73, p < 0.001) (Table S4). In low (Z= 4.88, p < 0.001) and intermediate (Z= 5.96, p < 0.001) light conditions dogs show more restive behaviour, while they were less restive when visibility of the surroundings was low from their site of resting (Z= -3.63, p < 0.001). Dogs were less relaxed in the resting site during the pup–rearing season (Z= -4.59, p < 0.001). During noon (Z= 5.49, p < 0.001) they were found to be more restive (Table S4).

**Resting proportion:**

The proportion of dogs resting at a time within a group is affected by temperature, light condition, position in the territory, season of the year and session of the day. The tendency of dogs to rest increases in high temperature (Z= 2.86, p < 0.001) and moderate light (Z= 2.57, p = 0.01), and decreases when the dogs rest in the intermediate position of the territory (Z= -2.23, p = 0.02) (Table S5). During the pre-mating and mating seasons (Z= -4.98, p < 0.001) resting tends to reduce, and the percentage of dogs resting at a time tend to increase during the morning (Z= 2.96, p < 0.001) (Table S5).

**Underlying patterns of site selection:**

We used only those individuals that were sighted at least six times during our entire observation period for this analysis, which resulted in a sample size of 5571 sightings. Within this refined dataset, we identified 1195 unique combinations of resting site characteristics based on the twelve physical parameters. To analyze the distribution of these unique combinations of parameters that described the resting sites, we plotted their frequency (normalized by the number of sightings for each unique dog) against their ranks. Surprisingly, the distribution exhibited a clear Power-law nature (Figure 1a), described by the equation $P(r) = C * r^{(-\alpha)}$, where P(r) represents the frequency of combinations, r is the rank, and α is the scaling exponent. The log-log plot of this data showed a slope of α = 2.9 (Figure 1b), leading to the equation $P(r) = 0.27 * r^{-2.907}$.

Since the GLMM analysis had revealed seven parameters that significantly contribute to the selection of resting sites with respect to different resting behaviours, we repeated the above analysis for the unique combinations of these seven physical properties of the resting sites. Remarkably, we found a significant correlation between this subset and an ordinal model of behavioural categories (Table S4). Within this subset, we documented 534 unique combinations of resting site characteristics preferred by free-ranging dogs. The frequency distribution of these unique combinations also followed a Power-law distribution (Figure S2a). The log-log plot of the data points revealed a slope of α = 2.34 (Figure S2b), with the equation $P(r) = 0.47 * r^{-2.34}$.

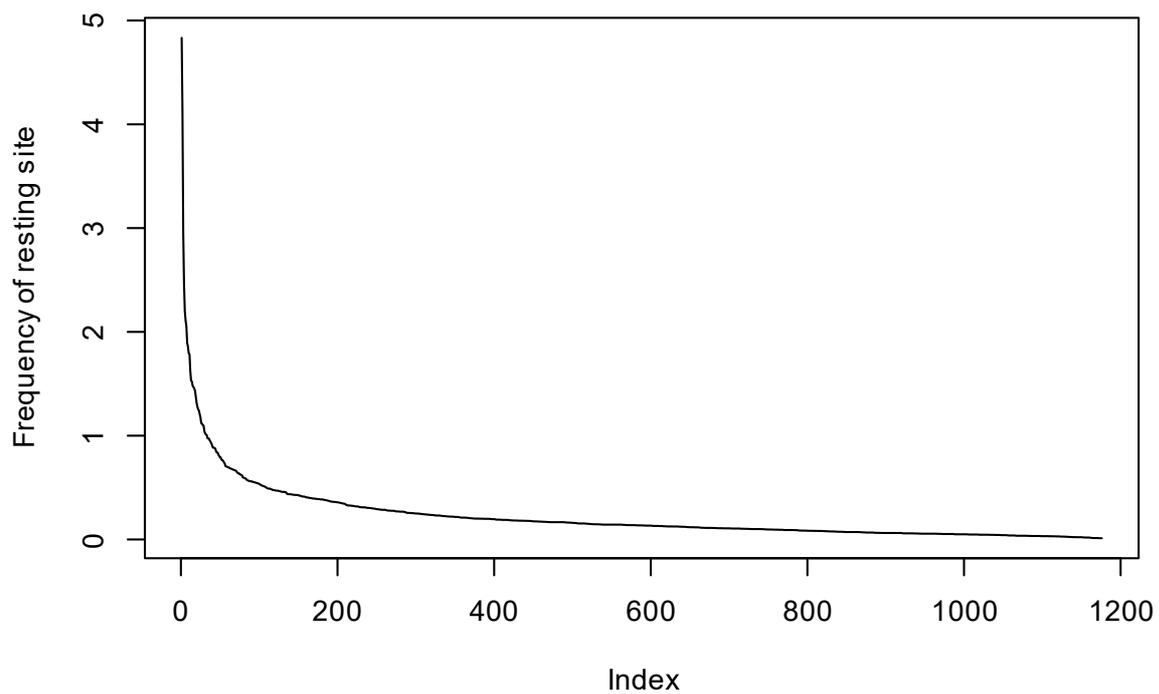

**Figure 1a.** Line plot showing the frequency distribution of unique combinations of twelve resting site parameters arranged according to their rank.

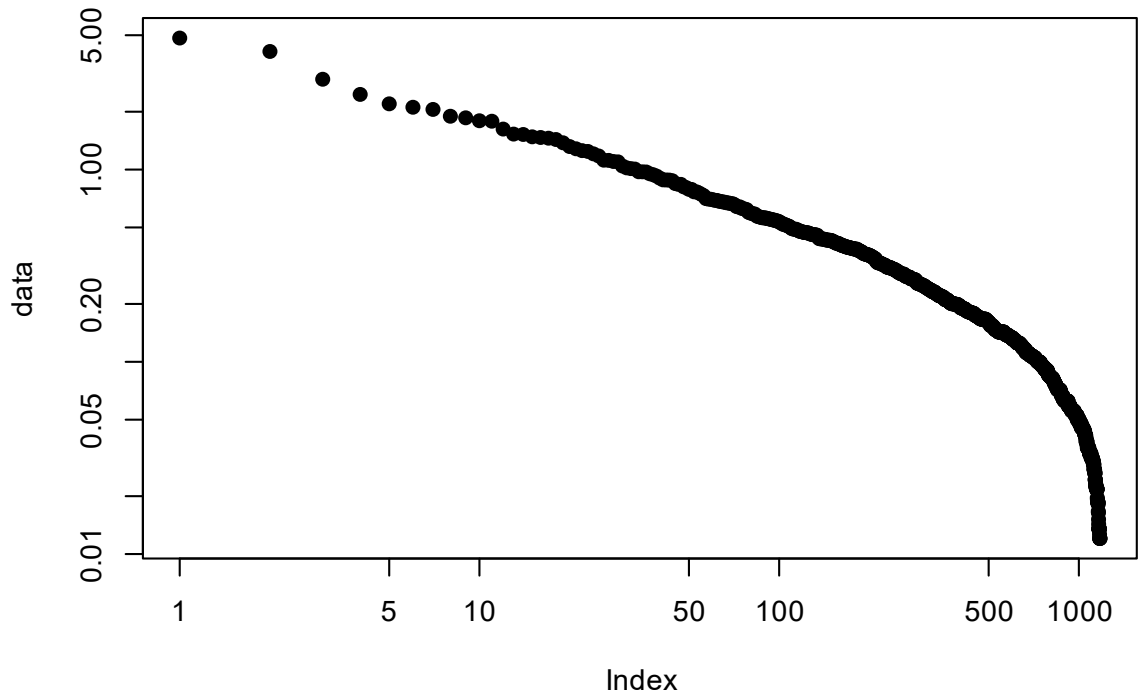

**Figure 1b.** The black dots represent the frequency-rank distribution of unique combinations of twelve resting site parameters plotted on a log-log scale.

**Understanding preference:**

For each of the 7 parameters identified, we calculated the proportion of resting for each behavioural category and for the overall data. Further, we identified the most preferred sub-parameter by comparing the rates at which these were chosen (Table 1).

**Comparison among seven parameters:**

| Category | DF | Disturbance | Light | PT | PSN | Shade | Visibility |
|---|---|---|---|---|---|---|---|
| Overall | <2 m | low | mod & high | center | beside road | (NS) | high |
| Extremely relaxed | <2 m | low | high | center | beside road | (NS) | high |
| Moderately relaxed | <2 m | (NS) | high | center | beside road | (NS) | high |
| Slightly relaxed | <2 m | high | mod & high | center | human related | yes | high |

**Table 1**: The key variables affecting different behavioural categories and the contexts in which they occur are summarized in this table. In terms of conduct, there are four categories: "Overall," "Extremely relaxed," "Moderately relaxed," and "Slightly relaxed." Key elements and contextual characteristics are described for each category to clarify the particular circumstances in which these actions are seen. DF signifies the distance from the food; PT implies position of the territory and PSN reflects the Place of resting.

Our findings indicate that free-ranging dogs exhibit a distinct preference for resting within a 2-meter radius of the food source. Their resting choices are closely linked to low disturbance levels, irrespective of whether the light conditions are moderate or high. Moreover, they tend to favor central areas within their territory. Their preference for resting locations aligns with road sides, where visibility is high, and interestingly, the presence or absence of shade does not appear to impact their resting decisions. Dogs selecting such sites are extremely relaxed, or perhaps, when they intend to rest well, they choose such locations. Similarly, dogs categorized as moderately relaxed follow the same trend as the overall pattern, with the notable exception of disturbance levels, which seem to have no effect on their resting.

In contrast, our analysis reveals a distinctive pattern for dogs exhibiting slightly relaxed behaviour. They predominantly opt to rest within 2 meters of food sources, even in areas characterized by high disturbance levels. Additionally, they appear to prefer human-related areas for resting. This choice is consistent with their inclination towards both moderate and high light conditions, along with a preference for resting in the presence of shade. Notably, their resting locations also coincide with areas offering high visibility. For further insights and detailed information on these patterns, refer to the Extended Supplementary Material (Table S6a-S6d).

Analyzing the resting patterns of dogs across various conditions has unveiled pronounced preferences for specific parameters. Irrespective of the level of resting, dogs always preferred to rest at a distance of less than 2 meters from the food source, favored the central position within their territory, and selected sites that offered them a high visibility of their surroundings.

To delve deeper into these preferences, we conducted a detailed comparison of the three most favored variables—distance from food, position within the territory, and visibility. Additionally, we examined the combinations of these two variables to gain a more comprehensive understanding of these preferences (Table 2). For an in-depth exploration of these analyses, refer to the Extended Supplementary Material (Table S7 & S8), which provides a detailed overview of the summary statistics.

| Category | Overall | Slightly relaxed | Moderately relaxed | Extremly relaxed |
|---|---|---|---|---|
| Single | df_<2m & pt_centre | df_<2m | df_<2m & pt_centre | pt_centre |
| Double | df_< 2m & pt_centre | df_< 2m & pt_centre | df_< 2m & pt_centre | df_< 2m & pt_centre |

| | | | pt_centre & visi_high | |
|---|---|---|---|---|

**Table 2:** Comparative analysis of preferred resting site parameters. This table provides a comprehensive comparison of the overall preferences for resting, considering the three most favored variables and their combinations. The variables are represented as follows: 'df_<2m': Denotes resting at a distance within 2 meters from the food source. 'pt_center': Signifies the preference for the center position within the territory. 'visi_high': Indicates the inclination for high visibility conditions at the resting site. This table elucidates the interplay of these variables and their joint influence on the resting preferences of the subjects. For a detailed breakdown of the findings, please refer to (Table S7 & S8).

In the overall condition, both resting within 2 meters from the food source (df_<2m) and occupying the center position of the territory (pt_center) emerged as favored choices. A similar trend was observed in dogs displaying moderately relaxed behaviour. However, among extremely relaxed dogs, the center position of the territory (pt_center) took precedence, while within slightly relaxed dogs, resting within 2 meters of the food source (df_<2m) was the most preferred option (Table S7).

Our investigation extended to analyzing the combinations of these variables. The combination 'df_<2m & pt_center' emerged as the most preferred choice for resting among free-ranging dogs. Remarkably, this pattern persisted in both extremely relaxed and slightly relaxed conditions among these dogs. Notably, in dogs with a moderately relaxed disposition, both 'df_<2m & pt_center' and 'pt_center & visibility_high' held equal preference. For a comprehensive overview of these findings, kindly refer to our analysis in Table S8.

Finally, in order to check if the combination of these three "preferred" (df: <2m_pt: center_and visibility: high_) parameters occurred in the whole data more than by chance alone, we estimated drew random combinations of three parameters compared the frequencies at which these were selected with the total frequency (normalized) at which sites with the three preferred sub-parameters had been selected. There were a total of 1666 sites which had the combination of the three preferred parameters, while the next highest combination contributed to less than 50% of this number. The preferred combination contributed to over 30% of the sites in the data (Figure 2). This suggested a strong underlying preference explained by the three parameters identified through our analysis.

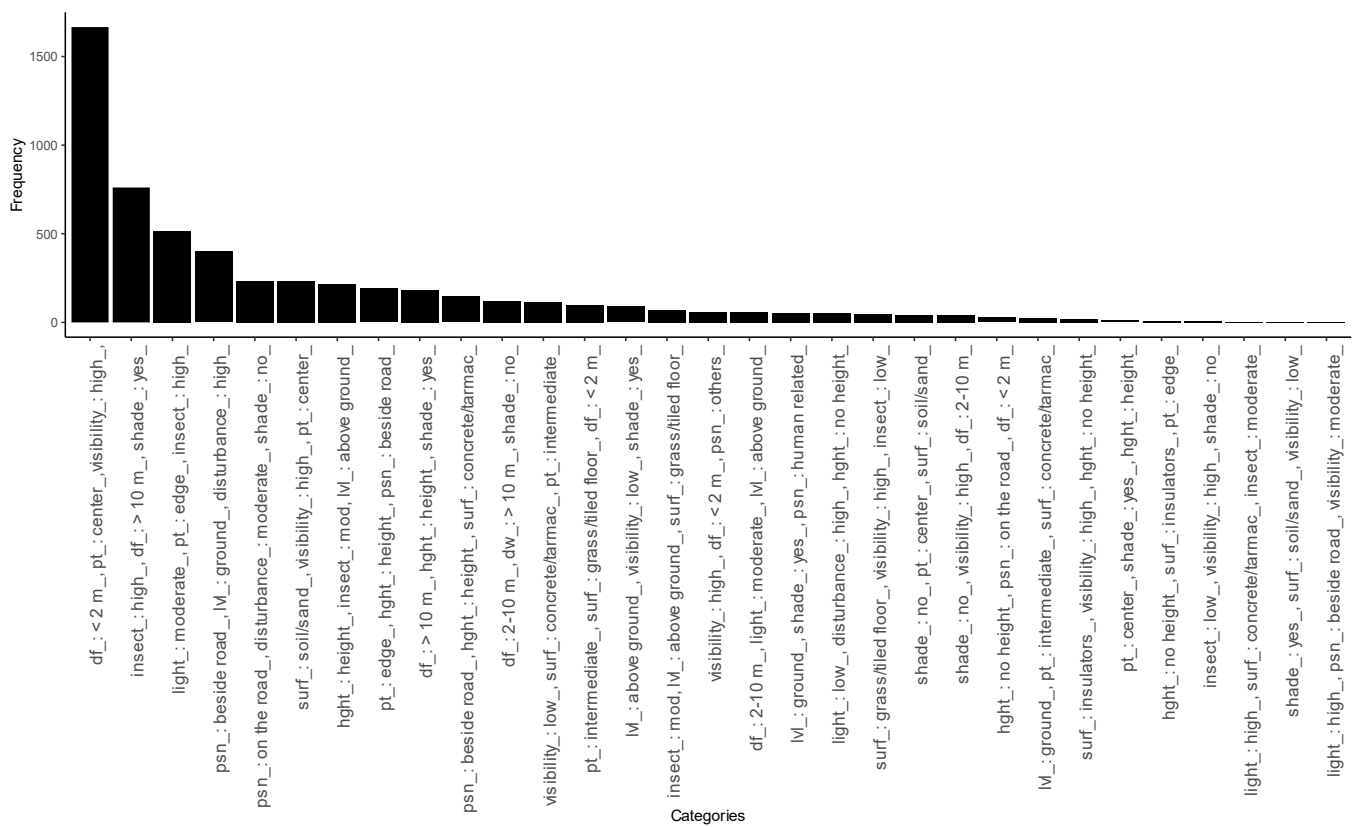

**Figure 2**. Describing the comparison of preferred (df: <2m_, pt: center_, visibility: high_) and 30 random combination of three parameters of resting. Here, 'df' refers to distance from food; dw: distance of water; pt: position of the territory; lvl: level of resting; hght: hight of resting; psn: position of resting site; surf: surface of resting.

**Discussion:**

We embarked on this study to understand how free-ranging dogs select sites to rest, and what characteristics of the resting sites might be responsible for the quality of rest. As expected, we identified a multitude of factors that affected resting behaviour. These were the distance of resources from the resting site, anthropogenic disturbance, position of the resting site within the territory, resting surface, shade, temperature, light condition of the resting site, visibility from the resting site and season of the year. However, it was also evident that the impact of the parameters on the quality of rest and the strength of the influence of various parameters on resting behaviour were varied. The frequency-rank distribution of the combination of resting site parameters followed a Power law distribution.

A power law is a mathematical pattern that describes a relationship where a change in one quantity is proportional to a power of another quantity. In a power law, smaller values of one variable are more common, while larger values are rarer, leading to a distribution with a heavy tail (Ulrich et al., 2010). In ecological contexts, power law distributions are often observed in various natural phenomena, such as species abundance distributions, connections between nodes in food webs and social network, the distribution of city sizes, the sizes of earthquakes, and even the distribution of wealth in societies (Clauset et al., 2009; Dunne et al., 2002). The occurrence of a power law distribution in resting site characteristics among free-ranging dogs reveals there are underlying biological drivers that explain our observations.

Though seven physical parameters influenced resting behaviour, for three of these (distance from resources, position in the territory and visibility), the dogs were very particular in their choice of a sub-parameter. Of the 189 possible combinations considering a random selection of three parameters in which only one of the sub-parameters is considered, the combination of the three preferred sub-parameters explained 30% of the data. In fact, a combination of only two such sub-parameters, namely, <2 m from food source and core position in the territory, could explain much of the preferences observed. Nevertheless, we discuss the various preferences of the dogs for resting site selection, and their implications in the ecology of the free-ranging dogs.

Resting within two metres of the food source, within the core area of the territory, and at a place that offered good visibility of the surroundings, was found to be of utmost importance to the dogs. Free-ranging dogs are primarily scavengers, and their food sources include garbage bins, open dumps, eateries, meat and fish shops in markets and households that give them food regularly (Majumder et al., 2014). We observed that they typically prefer to rest close to resources, but tend to be more restive when they are in the core of the territory, as opposed to when they are near the edge. It is likely that when they are resting close to a food source, they instinctively remain alert to guard the food source against intruders, which supports the "resource defense" and "foraging efficiency" hypotheses. This also agrees well with the finding that the dogs are less restive in areas of high anthropogenic disturbance, and when visibility of their surroundings is low. The core area of the territory is more "safe" for the dogs, as opposed to the edges and intermediate zones.

Thermoregulation is the method through which animals maintain a steady internal body temperature in the face of changing environmental conditions (Smith et al., 2007). As proposed by the thermoregulatory hypothesis and observed in some species, thermoregulation is a key priority for choosing resting or sleeping sites (Ellison et al., 2019; Podgórski et al., 2008; Smith et al., 2007). Moreover, dogs have a full covering of fur that can serve as insulation in cold weather but can be stressful in high temperature conditions, which leads them to find cool and shaded spaces for resting during the daytime, especially in the summer. In addition, dogs use panting to regulate body temperature; as they pant, moisture is drained from their respiratory system, which aids in heat dissipation (Crawford & Mechanical, 1962). Noon is typically the hottest part of the day, and since the dogs tend to prefer shady spaces and are more restive when the temperature is high, they are also more restive during the noon.

Canids are highly visual creatures that rely on their eyesight for several functions including hunting, social relationships and spatial navigation (Janko et al., 2012; Mech, 1989). Poor visibility conditions, such as dimly lit places or man-made structures with low visibility, might impair a dog's ability to perceive possible risks such as competitors and threatening humans. The presence of high anthropogenic disturbance is a very important source of stress for the dogs, and thus they might be more anxious and vigilant while resting in spaces with a high level of disturbance. When coupled with low visibility of the surroundings, even moderate levels of disturbance cause the dogs to be less restive. Similarly, humans can be a threat to dogs, and hence, they are less likely to be completely relaxed when they are resting in the vicinity of humans.

The presence of pups typically makes the mothers, and sometimes other group members more vigilant. They have to watch out for possible threats and the mothers spend a large part of their time in maternal care (Paul et al., 2017). Hence, even if she is resting, she might remain vigilant and guard the pups (Paul et al., 2017). Sometimes, other adult females and males in the group also guard pups and show alloparental care (Paul & Bhadra, 2018). This is likely to reduce the overall restive nature of the adults during the pup rearing season.

In addition to temperature, light conditions and position of the dog within the territory, the time of the year and of the day also contributed to the percentage of dogs of a group resting. Though the findings seem to counter some of the findings in relation to resting behaviour, we need to take a closer look at these results to understand their implications. While the earlier analysis reported the conditions under which the dogs are most relaxed (more restive), showing the highest level of energy conservation, as individuals, this analysis considers the group as the unit. Hence, while an individual resting dog can be more restive in the core area, when the group is present in the intermediate zone of the territory, the proportion of dogs showing some resting behaviour is higher than when they are in the edge (most dogs are vigilant) or the core (a lower proportion of dogs resting, but being more restive).

In the pre-mating and mating seasons, mostly adults are present in the population, and they are focused on finding mates. Hence, the tendency of dogs to rest is lower than in the pup-rearing season. Canids are nocturnal animals, and though dogs seem to have adapted to a more generalist lifestyle, living around humans (Jackson et al., 2014; Sen Majumder et al., 2014), more members of the group tend to rest at the same time during the morning as compared to the evening, when their territorial interactions increase and they tend to be more vigilant. Moreover, as reported earlier, they prefer to avoid direct sunlight and tend to be more restive when the ambient temperature increases.

This study began as an exploration, much of which was conducted during the pandemic, and was restricted by the regulations for movement on the streets. Nevertheless, we obtained data spread over different areas of West Bengal, India, following groups of dogs for over a year at a time. This gave us insights into the seasonal variation in resting site preferences, and provided a large pool of observations for analysing patterns in resting site selection. Our study revealed a multifaceted interplay of resting preferences among free-ranging dogs implying a flexible decision-making process they employ when picking resting spots. These findings help us understand the ecological adaptability, territorial tendencies, and responses to varied environmental conditions, of the free-ranging dogs. This research has ramifications for urban ecology, animal behaviour, and the cohabitation of free-roaming canines within human-dominated habitats. We could use insights from this study to design for inclusive and sustainable urban spaces that allow for co-habitation of the free-ranging dogs with humans, especially in countries of the Global South, where dog-human conflict is on the rise.

**Funding**

Sourabh Biswas was supported by a fellowship from the University Grants Commission, India. This work was funded by IISER Kolkata.

**Author contributions**

SB carried out the field work in IISER Kolkata, Gayeshpur and Sodepur. KG carried out field work in Burdawan. KS carried out field work in Banswara. SB and KG curated the data, SB carried out the data analysis. AB conceived the idea, supervised the work and co-wrote the paper with SB.

**References:**

Anderson, J. R. (2000). Sleep-related behavioural adaptations in free-ranging anthropoid primates. In *Sleep Medicine Reviews* (Vol. 4, Issue 4, pp. 355–373). W.B. Saunders Ltd. https://doi.org/10.1053/smrv.2000.0105

Atkins, D. L., by D Martin, E. R., Doyle, G. A., Walker, A. C., & David John Chivers Karger, B. S. (n.d.). The Museum, Oklahoma State University THE SIAMANG IN MALAYA. A Field Study of a Primate in Tropical Rain Forest. In *Conributions to Primatology* (Vol. 4). http://www.journals.uchicago.edu/t-and-c

Banerjee, A., & Bhadra, A. (2022). Time-activity budget of urban-adapted free-ranging dogs. *Acta Ethologica*, *25*(1), 33–42. https://doi.org/10.1007/s10211-021-00379-6

Brooks, M. E. ;, Kristensen, K. ;, Van Benthem, K. J. ;, Magnusson, A. ;, Berg, C. W. ;, Nielsen, A. ;, Skaug, H. J. ;, Machler, M. ;, Bolker, B. M., Brooks, M. E., Kristensen, K., Van Benthem, K. J., Magnusson, A., Berg, C. W., Nielsen, A., Skaug, H. J., & Mächler, M. (n.d.). *ETH Library glmmTMB balances speed and flexibility among packages for zero-inflated generalized linear mixed modeling Rights / license: Creative Commons Attribution 4.0 International glmmTMB Balances Speed and Flexibility Among Packages for Zero-inflated Generalized Linear Mixed Modeling*. https://doi.org/10.3929/ethz-b-000240890

Brown, D. D., Montgomery, R. A., Millspaugh, J. J., Jansen, P. A., Garzon-Lopez, C. X., & Kays, R. (2014). Selection and spatial arrangement of rest sites within northern tamandua home ranges. *Journal of Zoology*, *293*(3), 160–170. https://doi.org/10.1111/jzo.12131

Chapman, C. A., Chapman, L. J., & Melaughlin, R. L. (1989). Multiple central place foraging by spider monkeys: travel consequences of using many sleeping sites. In *Oecologia* (Vol. 79).

Clauset, A., Shalizi, C. R., & Newman, M. E. J. (2009). Power-law distributions in empirical data. In *SIAM Review* (Vol. 51, Issue 4, pp. 661–703). https://doi.org/10.1137/070710111

Coppinger, B., Cannistraci, R. A., Karaman, F., Kyle, S. C., Hobson, E. A., Freeberg, T. M., & Hay, J. F. (2017). Studying audience effects in animals: what we can learn from human language research. In *Animal Behaviour* (Vol. 124, pp. 161–165). Academic Press. https://doi.org/10.1016/j.anbehav.2016.12.020

Crawford, E. C., & Mechanical, J. R. (1962). *Mechanical aspects of panting in dogs'*. www.physiology.org/journal/jappl

Dunne, J. A., Williams, R. J., & Martinez, N. D. (n.d.). *Food-web structure and network theory: The role of connectance and size*. www.pnas.orgcgidoi10.1073pnas.192407699

Ellison, G., Wolfenden, A., Kahana, L., Kisingo, A., Jamieson, J., Jones, M., & Bettridge, C. M. (2019). Sleeping Site Selection in the Nocturnal Northern Lesser Galago (Galago senegalensis) Supports Antipredator and Thermoregulatory Hypotheses. *International Journal of Primatology*, *40*(2), 276–296. https://doi.org/10.1007/s10764-019-00085-y

Feilen, K. L., & Marshall, A. J. (2017). Multiple Ecological Factors Influence the Location of Proboscis Monkey (Nasalis larvatus) Sleeping Sites in West Kalimantan, Indonesia. *International Journal of Primatology*, *38*(3), 448–465. https://doi.org/10.1007/s10764-017-9953-1

Fisher, R. J., & Wiebe, K. L. (2006). Nest site attributes and temporal patterns of northern flicker nest loss: Effects of predation and competition. *Oecologia*, *147*(4), 744–753. https://doi.org/10.1007/s00442-005-0310-2

Gillespie, C. S. (2015). *Journal of Statistical Software Fitting Heavy Tailed Distributions: The poweRlaw Package*. http://www.jstatsoft.org/

Haubo, R., Christensen, B., & Maintainer, ]. (2022). *Package "ordinal" Title Regression Models for Ordinal Data*.

HENNER, C. M., CHAMBERLAIN, M. J., LEOPOLD, B. D., & BURGER, L. W. (2004). A MULTI-RESOLUTION ASSESSMENT OF RACCOON DEN SELECTION. *Journal of Wildlife Management*, *68*(1), 179–187. https://doi.org/10.2193/0022-541x(2004)068[0179:amaord]2.0.co;2

Herr, J., Schley, L., Engel, E., & Roper, T. J. (2010). Den preferences and denning behaviour in urban stone martens (Martes foina). *Mammalian Biology*, *75*(2), 138–145. https://doi.org/10.1016/j.mambio.2008.12.002

Jackson, C. R., John Power, R., Groom, R. J., Masenga, E. H., Mjingo, E. E., Fyumagwa, R. D., Røskaft, E., & Davies-Mostert, H. (2014). Heading for the hills: Risk avoidance drives den site selection in African wild dogs. *PLoS ONE*, *9*(6). https://doi.org/10.1371/journal.pone.0099686

Janko, C., Schröder, W., Linke, S., & König, A. (2012). Space use and resting site selection of red foxes (Vulpes vulpes) living near villages and small towns in Southern Germany. *Acta Theriologica*, *57*(3), 245–250. https://doi.org/10.1007/s13364-012-0074-0

Jucá, T., Boyle, S., Cavalcanti, G., Cavalcante, T., Tomanek, P., Clemente, S., De Oliveira, T., & Barnett, A. A. (2020). Being hunted high and low: do differences in nocturnal sleeping and diurnal resting sites of howler monkeys (Alouatta nigerrima and Alouatta discolor) reflect safety from attack by different types of predator? In *Biological Journal of the Linnean Society* (Vol. 131). https://academic.oup.com/biolinnean/article/131/1/203/5887841

Lowry, H., Lill, A., & Wong, B. B. M. (2013). Behavioural responses of wildlife to urban environments. *Biological Reviews*, *88*(3), 537–549. https://doi.org/10.1111/brv.12012

Lüdecke, D., Ben-Shachar, M., Patil, I., Waggoner, P., & Makowski, D. (2021). performance: An R Package for Assessment, Comparison and Testing of Statistical Models. *Journal of Open Source Software*, *6*(60), 3139. https://doi.org/10.21105/joss.03139

Lutermann, H., Verburgt, L., & Rendigs, A. (2010). Resting and nesting in a small mammal: Sleeping sites as a limiting resource for female grey mouse lemurs. *Animal Behaviour*, *79*(6), 1211–1219. https://doi.org/10.1016/j.anbehav.2010.02.017

Majumder, S. Sen, Bhadra, A., Ghosh, A., Mitra, S., Bhattacharjee, D., Chatterjee, J., Nandi, A. K., & Bhadra, A. (2014). To be or not to be social: Foraging associations of free-ranging dogs in an urban ecosystem. *Acta Ethologica*, *17*(1), 1–8. https://doi.org/10.1007/s10211-013-0158-0

McKinney, M. L. (2006). Urbanization as a major cause of biotic homogenization. *Biological Conservation*, *127*(3), 247–260. https://doi.org/10.1016/j.biocon.2005.09.005

Mckinney, M. L. (2008). *Effects of urbanization on species richness : A review of plants and animals*. 161–176. https://doi.org/10.1007/s11252-007-0045-4

Mech, L. D. (1989). Wolf Population Survival in an Area of High Road Density. In *Naturalist* (Vol. 121, Issue 2).

Meriggi, A., Rosa, P., Brangi, A., & MATTEUCCI Meriggi, C. A. (n.d.). *Habitat use and diet of the wolf in northern Italy*.

Nunn, C. L., & Heymann, E. W. (2005). Malaria infection and host behaviour: A comparative study of Neotropical primates. *Behavioural Ecology and Sociobiology*, *59*(1), 30–37. https://doi.org/10.1007/s00265-005-0005-z

Paul, M., & Bhadra, A. (2018). The great Indian joint families of free-ranging dogs. *PLoS ONE*, *13*(5). https://doi.org/10.1371/journal.pone.0197328

Paul, M., Majumder, S. Sen, & Bhadra, A. (2014). Grandmotherly care: A case study in Indian free-ranging dogs. *Journal of Ethology*, *32*(2), 75–82. https://doi.org/10.1007/s10164-014-0396-2

Paul, M., Sau, S., Nandi, A. K., & Bhadra, A. (2017). Clever mothers balance time and effort in parental care: A study on free-ranging dogs. *Royal Society Open Science*, *4*(1). https://doi.org/10.1098/rsos.160583

Podgórski, T., Schmidt, K., Kowalczyk, R., & Gulczyńska, A. (2008). Microhabitat selection by Eurasian lynx and its implications for species conservation. *Acta Theriologica*, *53*(2), 97–110. https://doi.org/10.1007/BF03194243

Purcell, K. L., Mazzoni, A. K., Mori, S. R., & Boroski, B. B. (2009). Resting structures and resting habitat of fishers in the southern Sierra Nevada, California. *Forest Ecology and Management*, *258*(12), 2696–2706. https://doi.org/10.1016/j.foreco.2009.09.041

Schneider, T. C., Kowalczyk, R., & Köhler, M. (2013). Resting site selection by large herbivores - The case of European bison (Bison bonasus) in Białowieża Primeval Forest. *Mammalian Biology*, *78*(6), 438–445. https://doi.org/10.1016/j.mambio.2013.06.002

Sen Majumder, S., Chatterjee, A., & Bhadra, A. (2014). *A dog's day with humans – time activity budget of free-ranging dogs in India* (Vol. 106, Issue 6). https://www.jstor.org/stable/24102275

Sen Majumder, S., Paul, M., Sau, S., & Bhadra, A. (2016). Denning habits of free-ranging dogs reveal preference for human proximity. *Scientific Reports*, *6*. https://doi.org/10.1038/srep32014

Smith, A. C., Knogge, C., Huck, M., Löttker, P., Buchanan-Smith, H. M., & Heymann, E. W. (2007). Long-term patterns of sleeping site use in wild saddleback (Saguinus fuscicollis) and mustached tamarins (S. mystax): Effects of foraging, thermoregulation, predation, and resource defense constraints. *American Journal of Physical Anthropology*, *134*(3), 340–353. https://doi.org/10.1002/ajpa.20676

Society, W. (2017). *Winter Resting Site Ecology of Marten in the Central Rocky Mountains Author ( s ): Steven W . Buskirk , Steven C . Forrest , Martin G . Raphael and Henry J . Harlow Published by : Wiley on behalf of the Wildlife Society Stable URL : http://www.jstor.org/s*. *53*(1), 191–196.

Taylor, S. L., & Buskirk, S. W. (1994). Forest microenvironments and resting energetics of the American marten Martes Americana. *Ecography*, *17*(3), 249–256. https://doi.org/10.1111/j.1600-0587.1994.tb00100.x

Vanak, A. T., Thaker, M., & Gompper, M. E. (2009). Experimental examination of behavioural interactions between free-ranging wild and domestic canids. *Behavioural Ecology and Sociobiology*, *64*(2), 279–287. https://doi.org/10.1007/s00265-009-0845-z

# Supplementary Material –

## Table S1. List of experimental sites

The table displays various sampling sites in the states of West Bengal, Maharashtra, and Rajasthan, India, along with their GPS locations, and specifies the type of habitat they represent.

All the sites in West Bengal are situated within the Indo-Gangetic plain, characterized by similar climatic conditions, receiving annual rainfall between mid-July and mid-September. The weather in this region can be described as a tropical wet and dry climate. In Maharashtra, the study sites in Pune are part of the Deccan Plateau in Western India. The climate here varies between tropical dry and wet, with hot summers, and the monsoon season typically lasting from June to October. On the other hand, Banswara, Rajasthan, experiences a tropical savanna climate, which is less extreme compared to the desert regions further north and northwest of Western India.

**Study area:**

The experiment was conducted between 2019 and 2022 in various parts of West Bengal, Maharashtra, and Rajasthan, India. In West Bengal, the study sites included the IISER-Kolkata campus, Bardawan, and the Sodepur area. We selected 10 groups from IISER-Kolkata, 17 groups from Bardawan, 12 groups from Gayeshpur, and 15 groups from the Sodepur area. The group sizes ranged from 2 to 19, and the number of groups differed across seasons.

| Location | Latitude | Longitude | Habitat type | Description |
|---|---|---|---|---|
| IISER-Kolkata | 22.96489 | 88.526614 | Semi-urban | Academic and research institute situated at Mohanpur, Nadia. It's situated in a semi-urban municipal area. |
| Gayeshpur | 22.958457 | 88.495431 | Semi-urban | It's a small town with a well-planned network of streets. |
| Bardawan | 23.252451 | 87.839693 | Semi-urban | The area is a semi-urban populated municipal area. |
| Sodepur | 22.705481 | 88.384817 | Urban | It's an urban area located close to the West Bengal state capital, Kolkata. |
| Pune | 18.545924 | 73.804247 | Urban | Pune is a metropolitan city in Maharashtra with an urban habitat. |
| Banswara | 23.544237 | 74.434291 | Semi-urban | The city of Banswara in Rajasthan is of semi-urban habitat. |

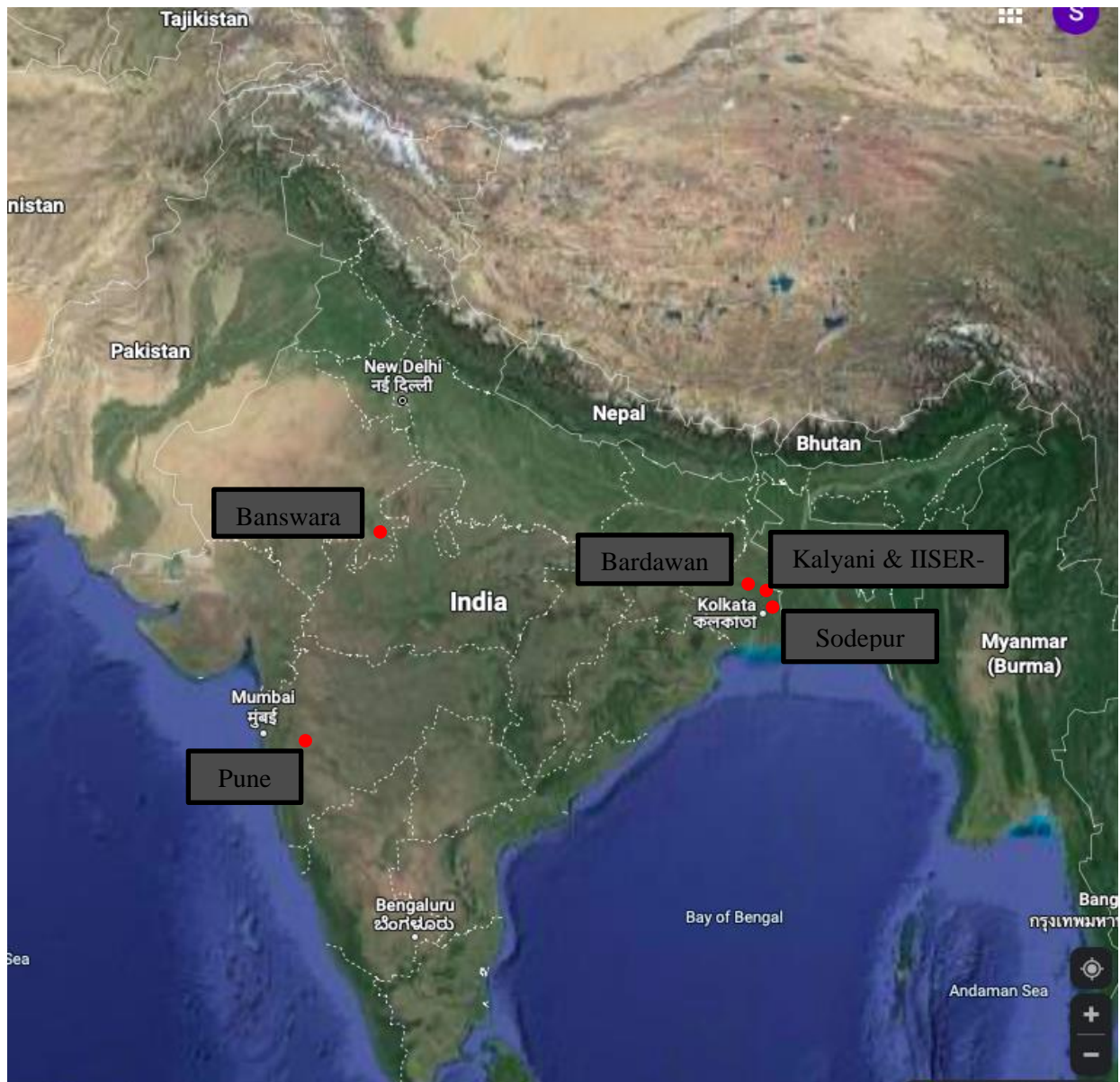

**Figure S1: Map of study sites**. The figure shows the various sites in which sampling was carried in the state of West Bengal, Maharashtra and Rajasthan, India.

**Table S2: Details of properties of resting site categories and sub-categories**

| Category | Sub-category | Description |
|---|---|---|
| **Distance from food** | 2-10 meter | When dogs rest in a spot located 2 to 10 meters away from their resource area, which is defined as the spot where dogs obtain food. This resource area can encompass various locations, including garbage dumps, household leftovers, chicken or fish shops, eateries, tea stalls, sweets or grocery shops. The study areas were previously surveyed to gain insight into the specific resource area. |
| | <2 meter | When dogs rest within a distance of less than 2 meters from their resource area. |
| | >10 meter | When dogs rest within a distance of more than 10 meters from their resource area. |
| **Distance from water** | 2-10 meter | When dogs rest at a location situated 2 to 10 meters away from the areas they use to drink water, such as ponds, drains, community tube wells, or water supply points where water is available and accessible to the dogs. The study areas were previously surveyed to gain insights into the specific locations where dogs drink water. |
| | <2 meter | When dogs rest within 2 meters of their water source area. |
| | >10 meter | When dogs rest more than 10 meters of their water source area. |
| **Disturbance (Foot fall)** | High | Disturbance or footfall of an area is determined by the average human footfall (including vehicles) at the resting site for one minute. High disturbance indicates footfall exceeding 40. |
| | Moderate | Moderate disturbance corresponds to a footfall ranging from 10 to 40. |
| | Low | Low disturbance indicates a footfall of less than 10. |
| **Position in territory** | Centre | The 'center of territory' refers to the core of the dog group's territory, approximately located at the middle point of their territory. We have defined this core area with a radius of 10-15 meters. The group territories of the dog(s) were previously surveyed to determine this core area. |
| | Intermediate | Intermediate position of territory refers to the middle area between the core and edge of a territory. |
| | Edge | Edge positions are either the boundaries between two adjacent dog groups or the outermost areas of a territory. Areas within a 5-meter distance from the territorial boundary are considered as edge positions. |
| **Preferred site** | Inside house | When a dog(s) is resting inside a house, the area inside the boundary walls of the house is also considered as inside the house. |
| | Entrance of a house | Entrance of a house refers to the gate or entryway of the house. |
| | Beside road | When a dog(s) is resting beside a road or footpath. |
| | On the road | When the dog(s) is resting on the road. |

| | | Road junction | When the dog(s) is resting at a junction between two or more roads. 'On the road' is used when the dog(s) is resting directly on a road. However, if the road includes any junction, 'Road junction' is used to describe the location, even if the dog(s) is resting on the road. |
|---|---|---|---|
| | | Under a vehicle | If dog(s) is resting under a vehicle. |
| | | Beside human | When a dog(s) rests in close proximity to a human, typically within a 2-meter radius, and the human remains stationary, we use the term 'beside human' to describe the location. This designation applies whether the dog(s) is resting inside a house, beside a road, or elsewhere. |
| | **Level** | Ground | We have used ground level, such as the road, as our reference point for surroundings. |
| | | Underground | Any height lower than the reference ground level is considered underground. This can include drains, dug holes, and similar features. |
| | | Above ground | Any height greater than the reference ground level is considered above ground. |
| | **Height** | No height | When a dog(s) is resting at the reference ground level, it is considered to be at no height. |
| | | >1 meter | If a dog(s) is resting at a level higher than the ground, and the height is greater than 1 meter, it is considered as '>1 meter'. |
| | | <1 meter | If a dog(s) is resting at a level other than the ground, and the height is less than 1 meter, it is considered as '<1 meter'. |

| Substrate or Surface | soil | If dog(s) is found resting on soil. |
|---|---|---|
| | Concrete/ tarmac | When a dog(s) is resting on a concrete, tarmac surface of a road, or a cemented area. |
| | Sand/ dust | If the dog(s) is resting on sand or dust, it is considered different from soil. Soil is considered a solid material, while sand and dust are finer and more porous. |
| | Grass | When a dog(s) is resting on a surface covered with grass. |
| | Leaf/ Garbage pile | If the dog(s) is resting in a leaf pile or a garbage pile. |
| | Tiled floor | When a dog(s) is resting on a tiled floor. |
| Shade | Under shade | When a dog(s) is resting in a spot that is shaded by a tree or a human-made roof, which can be either permanent or temporary. |
| | Open | If the resting spot has no shade and is therefore open. |
| Light | High | If the resting site has direct sunlight or any direct light sources. |
| | Moderate | When the resting site has light that is not as intense as 'High' but not as low as 'Low,' it may include shade. |
| | Low | Low light' means light intensity lower than 'moderate,' including dark conditions. |
| Visibility | High | If the resting area has a 360° view or is open on all four sides without any blockages, it is considered to have high visibility. |
| | Moderate | Visibility is considered moderate when the resting site has blockages from 1-2 sides. |
| | low | Visibility is considered low when the resting site has blockages on more than three sides. |

| **Insect index** | High | If the resting site has a high insect disturbance with more than approximately 30 individuals (mosquitoes, flies, ants and other arthopods). |
| --- | --- | --- |
| | Moderate | If the resting site has a moderate insect disturbance with approximately 5-30 individuals . |
| | Low | If the resting site has low insect disturbances with less than 5 individuals. |

**Table S3: Categorization of behaviour during resting in free-ranging dogs based on energy conservation.**

| Behaviour category | Behaviours |
| --- | --- |
| **Slightly relaxed** | Nursing, Sit or stand alert, Bark, Show affiliation, Watching, Solicit, Eat grass, Eat, Watching Dog, Snarl, Whoop, Lick, Removing and eating insects, Watching human, Tail wag, Feed, Beg, Tail shake, Flapping ears, Suckle, Squeal, Nibble dog, Whine, Growl, Digging, Howl, Dog gets beaten, Genital sniffing, Nape bite |
| **Moderately relaxed** | Gaze, Sitting, Groom, Lazing, Laid down, Lick self, Scratching, Sniff at, Yawning, Being Licked, Tongue flick, Sneeze, Nudge, Hang out tongue, Salivation, Sit up, Being Nibbled, Wake up, Nibble food/object, Pile Sleep, Stand, Roll, Chewing, Posture change |
| **Extremely relaxed** | Sleep, Curl up, Move while sleeping, Play, Snoring |

**Table S4: Effect of physical properties of resting site and weather parameters on resting behaviour of free-ranging dogs**

Ordinal mixed logistic regression (OMLR) analysis.

Model: (Behaviour category (Slightly relaxed/Moderately relaxed/ Extremely relaxed) ~ Distance form food (Df) + Disturbance + Light + Position of the territory (Pt) + Preferred site of resting (Psn) + Shade + Temperature + Visibility + Season + Session of the day, random factors (Dogs).

| Variables | Estimate | Std. Error | z value | Pr(>z) |
| --- | --- | --- | --- | --- |
| Df< 2 m | -0.4 | 0.07 | -5.54 | 3.06E-08 |
| Df> 10 m | -0.11 | 0.11 | -0.95 | 3.41E-01 |
| Disturbance_low | 0.6 | 0.08 | 7.49 | 6.87E-14 |
| Disturbance_moderate | 0.15 | 0.07 | 2.01 | 4.44E-02 |
| Light_low | 0.64 | 0.13 | 4.88 | 1.04E-06 |
| Light_moderate | 0.45 | 0.08 | 5.97 | 2.42E-09 |
| Pt_edge | -0.75 | 0.09 | -8.01 | 1.11E-15 |
| Pt_intermediate | -0.16 | 0.08 | -2.04 | 0.0413 |
| Psn_human related | -0.72 | 0.07 | -9.6 | < 2.22e-16 |
| Psn_on the road | -0.14 | 0.07 | -1.91 | 0.0562 |
| Psn_others | 0.12 | 0.18 | 0.68 | 0.4974 |
| Shade_yes | 0.61 | 0.06 | 10.03 | < 2.22e-16 |
| Temp | -0.12 | 0.01 | -14.73 | < 2.22e-16 |
| Visibility_low | -0.41 | 0.11 | -3.62 | 0.00029 |
| Visibility_moderate | 0.03 | 0.09 | 0.33 | 0.7451 |

| Variable | Estimate | Std. Error | z value | Pr(>z) |
|---|---|---|---|---|
| Season_premating-mating | 0.08 | 0.06 | 1.32 | 1.88E-01 |
| Season_pup-rearing | -0.66 | 0.14 | -4.59 | 4.40E-06 |
| Session_morning | 0.4 | 0.09 | 4.61 | 3.93E-06 |
| Session_night | -0.48 | 0.29 | -1.64 | 1.01E-01 |
| Session_noon | 0.55 | 0.1 | 5.5 | 3.84E-08 |

**Table S5: Effect of physical properties of resting site and weather parameters on overall resting propensity in free-ranging dog groups.**

Generalized Linear Mixed Model (GLMM):

Model: (Resting proportion of group ~ Distance form food (Df) + Distance form water (Dw) + Height of resting site (Hght) + Temperature (Temp) + Presence of Insect + Level of resting site + Light + Position of the territory (Pt) + Preferred site of resting (Psn) + Shade + Surface of resting (Surf) + Season + Session of the day, random factors (Dog group).

| Variable | Estimate | Std. Error | z value | Pr(>z) |
|---|---|---|---|---|
| (intercept) | -0.45 | 0.24 | -1.9 | 0.057 |
| Df> 10 m | -0.01 | 0.1 | -0.1 | 0.916 |
| Df2-10 m | -0.06 | 0.07 | -0.95 | 0.344 |
| Dw> 10 m | -0.01 | 0.09 | -0.16 | 0.876 |
| Dw2-10 m | 0.02 | 0.06 | 0.28 | 0.781 |
| Hght_no height | -0.04 | 0.13 | -0.26 | 0.793 |
| Temp | 0.02 | 0.01 | 2.86 | 0.004 ** |
| Insectmod_high | 0.08 | 0.05 | 1.5 | 0.133 |
| Insect_moderate | 0.16 | 0.11 | 1.46 | 0.145 |
| Level_ground | 0.16 | 0.13 | 1.2 | 0.228 |
| Level_under ground | 0.12 | 0.17 | 0.74 | 0.458 |
| Light_low | 0.09 | 0.08 | 1.18 | 0.237 |
| Light_moderate | 0.14 | 0.05 | 2.57 | 0.010 * |
| Pt_edge | 0.05 | 0.07 | 0.69 | 0.492 |
| Pt_intermediate | -0.15 | 0.07 | -2.24 | 0.025 * |
| Psn_human related | 0.11 | 0.06 | 1.87 | 0.062 . |
| Psn_on the road | 0.07 | 0.07 | 1 | 0.315 |
| Psn_others | 0.07 | 0.12 | 0.55 | 0.582 |
| Shade_yes | 0 | 0.05 | 0.09 | 0.929 |
| Surf_grass/tiled floor | 0.14 | 0.1 | 1.5 | 0.134 |
| Surf_insulators | -0.2 | 0.13 | -1.5 | 0.134 |
| Surf_soil/sand | 0.07 | 0.06 | 1.22 | 0.221 |
| Season_premating-mating | -0.25 | 0.05 | -4.99 | 6.13e-07 *** |
| Season_pup-rearing | -0.2 | 0.11 | -1.88 | 0.060 . |
| Session_morning | 0.19 | 0.06 | 2.97 | 0.003 ** |
| Session_night | -0.04 | 0.22 | -0.19 | 0.85 |
| Session_noon | -0.04 | 0.07 | -0.55 | 0.584 |

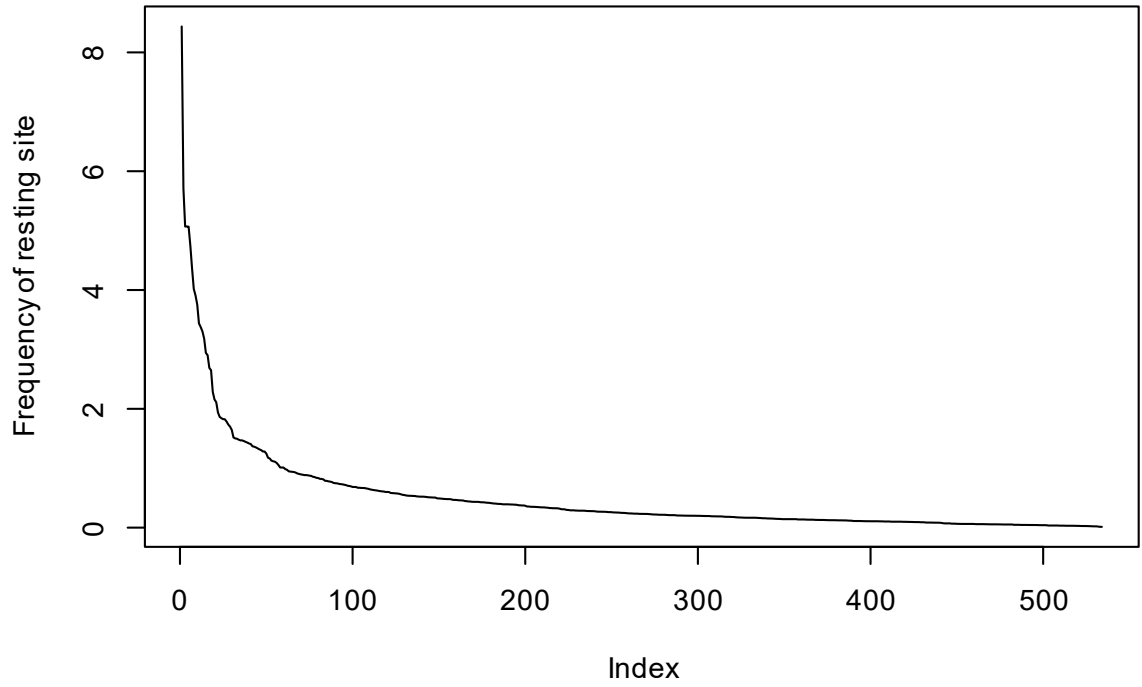

**Figure S2a:** Line plot showing the frequency distribution of unique combinations of seven resting site parameters arranged according to their rank.

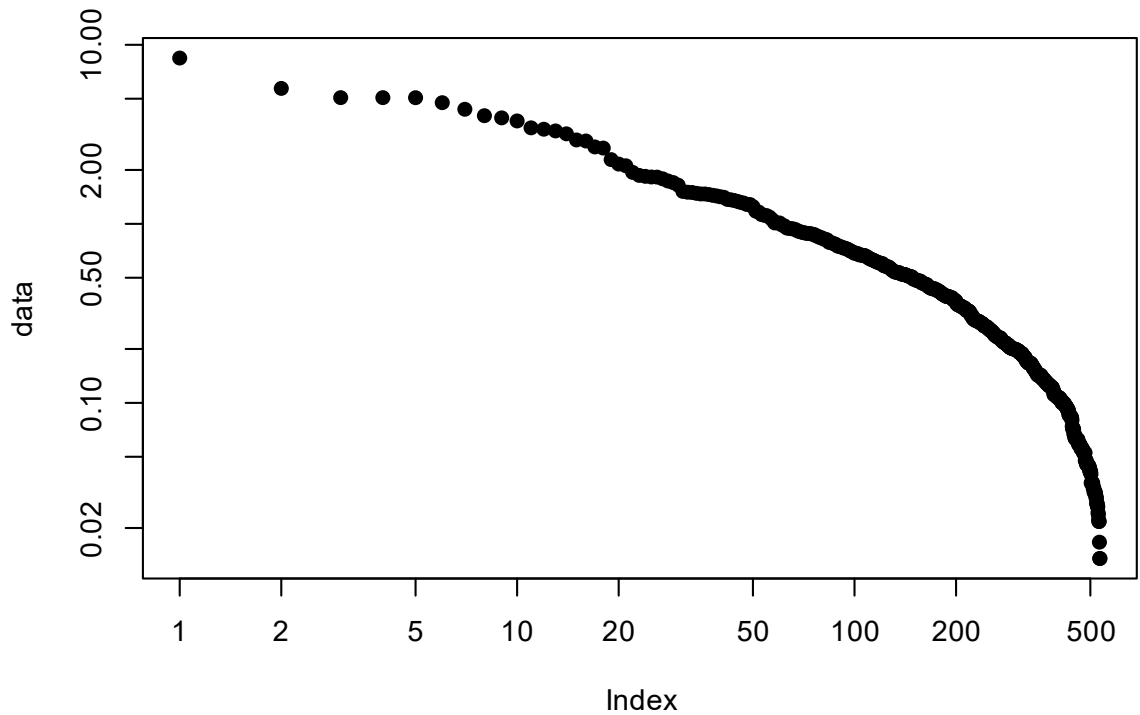

**Figure S2b:** The black dots represent the frequency-rank distribution of unique combinations of seven resting site parameters plotted on a log-log scale.

**Table. S6a: Comparison on physical parameters of resting site preferences in free-ranging dogs in overall condition.**

| Category | Mean ± SD | Kruskal-Wallis $\chi^2$ | df | p-value |
|---|---|---|---|---|
| Distance | | | | |
| < 2 m | 0.70 ± 0.26 | 164.11 | 2 | 0 |
| > 10 m | 0.29 ± 0.26 | | | |
| 2-10 m | 0.29 ± 0.23 | | | |
| Disturbance | | | | |
| High | 0.36 ± 0.23 | 10.05 | 2 | 0.01 |
| Low | 0.46 ± 0.30 | | | |
| Moderate | 0.36 ± 0.21 | | | |
| | | | | |
| Light | | | | |
| High | 0.43 ± 0.22 | 89.6 | 2 | 0 |
| Low | 0.24 ± 0.14 | | | |
| Moderate | 0.39 ± 0.21 | | | |
| | | | | |
| Territory position | | | | |
| Centre | 0.72 ± 0.30 | 174.03 | 2 | 0 |
| Edge | 0.24 ± 0.24 | | | |
| Intermediate | 0.28 ± 0.26 | | | |
| | | | | |
| Place of resting | | | | |
| Beside Road | 0.52 ± 0.28 | 133.83 | 3 | 0 |
| Human Related | 0.27 ± 0.18 | | | |
| On the Road | 0.31 ± 0.23 | | | |
| Others | 0.14 ± 0.18 | | | |
| | | | | |
| Shade | | | | |
| No | 0.52 ± 0.22 | 0.05 | 1 | 0.82 |
| Yes | 0.52 ± 0.24 | | | |

| Category | | | | |
|---|---|---|---|---|
| Visibility | | | | |
| High | 0.59 ± 0.22 | 232.56 | 2 | 0 |
| Low | 0.25 ± 0.16 | | | |
| Moderate | 0.23 ± 0.17 | | | |

**Table S6b: Comparison on physical parameters of resting site preferences in free-ranging dogs in extremely relaxed condition.**

| Category | Mean | SD | Kruskal-Wallis | df | p-value |
|---|---|---|---|---|---|
| Distance | | | | | |
| < 2 m | 0.71 | 0.26 | 53.86 | 2 | 0 |
| > 10 m | 0.3 | 0.38 | | | |
| 2-10 m | 0.33 | 0.23 | | | |
| Light | | | | | |
| High | 0.43 | 0.26 | 18.74 | 2 | 0 |
| Low | 0.27 | 0.19 | | | |
| Moderate | 0.4 | 0.21 | | | |
| Territory position | | | | | |
| Center | 0.82 | 0.25 | 78.07 | 2 | 0 |
| Edge | 0.24 | 0.2 | | | |
| Intermediate | 0.25 | 0.23 | | | |
| | | | | | |
| Disturbance | | | | | |
| High | 0.27 | 0.21 | 43.04 | 2 | 0 |
| Low | 0.56 | 0.27 | | | |
| Moderate | 0.36 | 0.23 | | | |
| | | | | | |
| Preferred site of resting | | | | | |
| Beside Road | 0.58 | 0.33 | 34.8 | 3 | 0.0001 |
| Human Related | 0.33 | 0.24 | | | |
| On the Road | 0.29 | 0.24 | | | |
| Others | 0.18 | 0.13 | | | |
| | | | | | |
| Visibility | | | | | |
| High | 0.58 | 0.26 | 58.41 | 2 | 0 |
| Low | 0.27 | 0.19 | | | |
| Moderate | 0.27 | 0.21 | | | |
| | | | | | |
| Shade | | | | | |
| No | 0.42 | 0.24 | 20.46 | 1 | 0 |
| Yes | 0.62 | 0.26 | | | |

**Table S6c: Comparison on physical parameters of resting site preferences in free-ranging dogs in moderately relaxed condition.**

| Category | Mean | SD | Kruskal-Wallis $c^2$ | df | p-value |
|---|---|---|---|---|---|
| **Distance** | | | | | |
| < 2 m | 0.76 | 0.23 | 131.93 | 2 | 0 |
| > 10 m | 0.25 | 0.18 | | | |
| 2-10 m | 0.23 | 0.16 | | | |
| ... | | | | | |
| **Light** | | | | | |
| High | 0.5 | 0.24 | 76.91 | 2 | 0 |
| Low | 0.2 | 0.11 | | | |
| Moderate | 0.36 | 0.23 | | | |
| ... | | | | | |
| **Territory position** | | | | | |
| Center | 0.74 | 0.3 | 84.76 | 2 | 0 |
| Edge | 0.31 | 0.26 | | | |
| Intermediate | 0.26 | 0.22 | | | |
| ... | | | | | |
| **Disturbance** | | | | | |
| High | 0.41 | 0.24 | 1.53 | 2 | 0.46 |
| Low | 0.47 | 0.31 | | | |
| Moderate | 0.39 | 0.2 | | | |
| ... | | | | | |
| **Place of resting** | | | | | |
| Beside Road | 0.59 | 0.27 | 34.8 | 3 | 0.0001 |
| Human Rel... | 0.26 | 0.18 | | | |
| On the Road | 0.29 | 0.21 | | | |
| Others | 0.17 | 0.16 | | | |
| ... | | | | | |
| **Visibility** | | | | | |
| High | 0.68 | 0.18 | 175.86 | 2 | 0 |
| Low | 0.21 | 0.12 | | | |
| Moderate | 0.17 | 0.14 | | | |
| ... | | | | | |
| **Shade** | | | | | |
| No | 0.52 | 0.24 | 0.69 | 1 | 0.41 |
| Yes | 0.49 | 0.24 | | | |

**Table S6d: Comparison on physical parameters of resting site preferences in free-ranging dogs slightly relaxed condition.**

| Category | Mean | SD | Kruskal-Wallis | df | p-value |
|---|---|---|---|---|---|
| Distance | | | | | |
| < 2 m | 0.71 | 0.26 | 53.86 | 2 | 0 |
| > 10 m | 0.3 | 0.38 | | | |
| 2-10 m | 0.33 | 0.23 | | | |
| Light | | | | | |
| High | 0.43 | 0.26 | 18.74 | 2 | 0 |
| Low | 0.27 | 0.19 | | | |
| Moderate | 0.4 | 0.21 | | | |
| | | | | | |
| Territory position | | | | | |
| Center | 0.82 | 0.25 | 78.07 | 2 | 0 |
| Edge | 0.24 | 0.2 | | | |
| Intermediate | 0.25 | 0.23 | | | |
| | | | | | |
| Disturbance | | | | | |
| High | 0.27 | 0.21 | 43.04 | 2 | 0 |
| Low | 0.56 | 0.27 | | | |
| Moderate | 0.36 | 0.23 | | | |
| | | | | | |
| Preferred site of resting | | | | | |
| Beside Road | 0.58 | 0.33 | 34.8 | 3 | 0.0001* |
| Human Related | 0.33 | 0.24 | | | |
| On the Road | 0.29 | 0.24 | | | |
| Others | 0.18 | 0.13 | | | |
| | | | | | |
| Visibility | | | | | |
| High | 0.58 | 0.26 | 58.41 | 2 | 0 |
| Low | 0.27 | 0.19 | | | |
| Moderate | 0.27 | 0.21 | | | |
| Shade | | | | | |
| No | 0.42 | 0.24 | 20.46 | 1 | 0.0000* |
| Yes | 0.62 | 0.26 | | | |

**Table S7: Single level comparison of resting site preferences in free-ranging dogs across four relaxation levels based on three physical parameters**

| Condition | Categories | Resting proportion (Mean ± SD) | Kruskal-Wallis $\chi^2$ | p-value | p value of post-hoc comparison vs. df_<2 m |
|---|---|---|---|---|---|
| Overall | df_<2 m | 0.70 ± 0.26 | 44 | 0.00 | N/A |
| | pt_centre | 0.72 ± 0.30 | | | 0.08; (<0.001, vs visibility_high) |
| | visibility_high | 0.59 ± 0.22 | | | <0.001 |
| Slightly Relaxed | df_<2 m | 0.91 ± 0.14 | 43.48 | 0.00 | N/A |
| | pt_centre | 0.71 ± 0.34 | | | 0.01; (<0.001, vs visibility_high) |
| | visibility_high | 0.50 ± 0.17 | | | <0.001 |
| Moderately Relaxed | df_<2 m | 0.76 ± 0.23 | 18.44 | 0.00 | N/A |
| | pt_centre | 0.74 ± 0.30 | | | >0.05; (<0.001, vs visibility_high) |
| | visibility_high | 0.68 ± 0.18 | | | 0.001 |
| Extremely Relaxed | df_<2 m | 0.71 ± 0.25 | 38.82 | 0.00 | N/A |
| | pt_centre | 0.82 ± 0.25 | | | 0.002; (<0.001, vs visibility_high) |
| | visibility_high | 0.58 ± 0.25 | | | 0.003 |

**Table S8: Comparison of resting site preferences at a combination of two in free-ranging dogs across four relaxation levels based on three physical parameters.**

| Condition | Categories | Resting Proportion (Mean ± SD) | Kruskal-Wallis $\chi^2$ | p-value | Post-hoc comparison |
|---|---|---|---|---|---|
| **Overall** | df_<2m & pt_centre | 0.52 ± 0.30 | 16.2 | 0.001 | N/A |
| | df_<2m & visib_high | 0.41 ± 0.21 | | | p < 0.001*, vs df_<2m & pt_centre |
| | pt_centre & visib_high | 0.46 ± 0.27 | | | p < 0.04* ; vs df_<2m & pt_centre (p = 0.09, vs df_<2m & visib_high) |
| **Slightly Relaxed** | df_<2m & pt_centre | 0.62 ± 0.32 | 22.8 | 0.001 | N/A |
| | df_<2m & visib_high | 0.42 ± 0.18 | | | p < 0.001; vs df_<2m & pt_centre (p = 0.21, vs pt_centre & visib_high) |
| | pt_centre & visib_high | 0.37 ± 0.21 | | | p = 0 vs df_<2m & pt_centre |
| **Moderately Relaxed** | df_<2m & pt_centre | 0.56 ± 0.29 | 6.57 | 0.019 | N/A |
| | df_<2m & visib_high | 0.48 ± 0.23 | | | p = 0.0193; vs df_<2m & pt_centre (p > 0.05, vs pt_centre & visib_high) |
| | pt_centre & visib_high | 0.53 ± 0.27 | | | p > 0.05, vs df_<2m & pt_centre |
| **Extremely Relaxed** | df_<2m & pt_centre | 0.63 ± 0.29 | 30.57 | 0.001 | N/A |
| | df_<2m & visib_high | 0.43 ± 0.25 | | | p < 0.001; vs df_<2m & pt_centre (p = 0.08, vs pt_centre & visib_high) |
| | pt_centre & visib_high | 0.50 ± 0.26 | | | p = 0.001, vs df_<2m & pt_centre |